# Low temperature optical properties of interstellar and circumstellar icy silicate grain analogues in the mid-infrared spectral region*

Alexey Potapov[1], Harald Mutschke[2], Phillip Seeber[1], Thomas Henning[3], Cornelia Jäger[1]

[1]*Laboratory Astrophysics Group of the Max Planck Institute for Astronomy at the Friedrich Schiller University Jena, Institute of Solid State Physics, Helmholtzweg 3, 07743 Jena, Germany, email: alexey.potapov@uni-jena.de*

[2]*Astrophysical Institute and University Observatory, Friedrich Schiller University, Schillergässchen 2-3, 07745 Jena, Germany*

[3]*Max Planck Institute for Astronomy, Königstuhl 17, D-69117 Heidelberg, Germany*

**Abstract**

Two different silicate/water ice mixtures representing laboratory analogues of interstellar and circumstellar icy grains were produced in the laboratory. For the first time, optical constants, the real and imaginary parts of the complex refractive index, of such silicate/water ice mixtures were experimentally determined in the mid-infrared spectral region at low temperatures. In addition, optical constants of pure water ice and pure silicates were derived in the laboratory. Two sets of constants were compared, namely, "measured" constants calculated from the transmission spectra of silicate/ice samples and "effective" constants calculated from the optical constants of pure silicates and pure water ice samples using different mixing rules (effective medium approaches). Differences between measured and effective constants show that a mixing (averaging) of optical constants of water ice and silicates for the determination of the optical properties of silicate/ice mixtures can lead to incorrect results. Also, it is shown that a part of water ice molecules is trapped in/on silicate grains and does not desorb up to 200 K. Our unique data are just in time with respect to the new challenging space mission, James Webb Space Telescope, which will be able to bring novel detailed information on interstellar and circumstellar grains, and suitable laboratory data are extremely important for the decoding of astronomical spectra.

1. **Introduction**

Dust grains play a central role in the physics and chemistry of practically all astrophysical environments. They influence the thermodynamic properties of the medium and provide a surface for very efficient chemical reactions responsible for the synthesis of a major part of important astronomical molecules. Interstellar and circumstellar dust grains are of great interest as building blocks of stars and planets and as a remnant of protoplanetary disks. Knowing the properties of grains, we can follow their pathways to larger astronomical bodies and trace back the history of planetary systems.

Dust grains are typically nanometre- to micrometre-sized aggregates of carbonaceous or siliceous particles (Henning 1997; Draine 2003). In cold cosmic environments, such as dense molecular clouds and outer parts of circumstellar shells, including envelopes of evolved stars, protoplanetary disks, and debris disks, dust grains are mixed with molecular ices. Observations and dedicated laboratory experiments have shown that the main constituent of interstellar and circumstellar ices is $H_2O$ with lower fractions of other volatile species (Allamandola et al. 1999; Burke & Brown 2010; van Dishoeck 2014; Boogert, Gerakines, & Whittet 2015).

Dust grains absorb and scatter stellar light and reemit the absorbed energy. To interpret the astronomical spectra of dust grains the spectral data on laboratory dust grain analogues are required. Optical properties of grains in different frequency regions are important for the modelling and understanding of the physics in astrophysical environments. The opacity of grains is the base for the estimation of important astrophysical parameters, such as dust temperatures, mass loss rates of evolved stars, and the total dust mass in circumstellar shells or molecular clouds.

The infrared range (wavelength from 700 nm to 1 mm) is one of the most important spectral regions with respect to dust grains because it brings information on vibrational modes of dust-building structural units and it is the range, mainly far infrared (wavelength from 0.1 mm to 1 mm), where the emission from interstellar dust is detected. Thus, infrared spectroscopy is the best astronomical and laboratory tool for studying the composition and properties of the cosmic dust and its laboratory analogues.

Optical properties of pure water ice, silicate and carbon grains have been intensively investigated in the laboratory (Hagen, Tielens, & Greenberg 1981; Kitta & Krätschmer 1983; Smith et al. 1994; Dorschner et al. 1995; Henning & Mutschke 1997; Jäger, Mutschke, & Henning 1998; Jäger et al. 2003; Curtis et al. 2005; Jäger et al. 2008; Mastrapa et al. 2008; Mastrapa et al. 2009; Allodi et al. 2014; Sabri et al. 2014; Reinert et al. 2015). The real astronomical grains in cold environments are mixtures of dust and ice, that is why it is a strong need to study experimentally optical properties of such mixtures, which are not yet known. One can try to obtain optical constants of dust/ice mixtures by mixing of known pure dust and ice constants using a number of mixing rules, as it was done in a number of works (Mukai & Mukai 1984; Maron & Maron 2008; Min et al. 2016), but the question is the reliability of such approaches.

The main purpose of this work is to present unique sets of experimental optical data for silicate/water ice mixtures in the mid-IR spectral region at low temperatures and to check the reliability of different mixing rules for production of optical data for such mixtures from the optical data of its components, silicates and water ice. The new optical constants of silicate/water ice mixtures published here can be used in models describing molecular clouds and circumstellar shells and in predictions of observable IR features.

## 2. Experimental procedure

Amorphous silicate $MgSiO_3$/water ice mixtures were produced in a device consisting of a laser ablation setup combined with a cryogenic chamber, which is able to condense solid carbonaceous and siliceous grains and gases at temperatures down to 8 K. The setup (without the cryostat) is described in detail elsewhere (Jäger, et al. 2008). The gas-phase deposition of nanometre-sized amorphous $MgSiO_3$ grains was achieved by pulsed laser ablation of a MgSi 1:1 target and subsequent condensation of the evaporated species in a quenching atmosphere of 4 mbar $O_2$. Condensed grains were extracted adiabatically from the ablation chamber through a nozzle into a second low pressure chamber ($10^{-3}$ mbar) to decouple the grains from the gas. A second extraction through a skimmer into a third chamber ($10^{-6}$ mbar) generated a particle beam that was directed into a fourth, cryogenic, chamber, where the grains were deposited onto a cold CsI substrate. Water was deposited simultaneously with the grains from a water reservoir through a leakage valve and a capillary tube. All depositions were performed in a high vacuum chamber with a base pressure of $9\times10^{-8}$ mbar at the temperature of the substrate of 8 K. Such relatively poor vacuum conditions due to the combination of a laser ablation system with the cryogenic chamber lead to a deposition of $CO_2$ from the chamber atmosphere. $CO_2$ stretching band is visible in the IR spectra recorded. However, the amount of $CO_2$ is small compared to the amount of the main deposits and cannot influence the optical properties discussed in this study. The deposition time was 20 minutes for all samples. The water deposition rate was a few tens of nm/min depending on the ice thickness. Such a low deposition rate corresponds to the formation of high-density amorphous water ice (Mastrapa, et al. 2009).

The thickness of the silicate grain deposit was measured by a quartz crystal resonator microbalance (sensitivity 0.1 nm) using known values for the deposit area and density. The particle beam was divided by an aperture to simultaneously deposit the grains on the microbalance and the substrate. Due to a slight tilt of the beam, the microbalance might be not completely covered. The uncertainty related to the determination of the area of the deposit on the microbalance is estimated to be about 5%. The thickness of the water matrix was estimated from the 3.1 μm water band area using the band strength of $2 \times 10^{-16}$ cm molecule$^{-1}$ (Hudgins et al. 1993) with an uncertainty of 2%. A 280 nm $MgSiO_3$ layer and a 640 nm $H_2O$ layer, and two silicate/water ice mixtures with the $MgSiO_3/H_2O$ mass ratios of 0.8 and 2.7 were deposited and studied. Additional measurements were also done for mixtures with the mass ratios of 0.3 and 1.5 to establish the mass ratio dependence of water trapped at 200 K. The mass ratio was calculated from the thicknesses of the deposits and densities of 1.1 g cm$^{-3}$ for high-density amorphous water ice (Narten 1976) and 2.5 g cm$^{-3}$ for amorphous silicates. For glassy silicates the density of 2.71 was measured (http://www.astro.uni-jena.de/Laboratory/OCDB/index.html). $MgSiO_3$ grains produced by laser ablation are characterized by less dense structures and show slightly lower values of 2.6 – 2.5. The thicknesses corresponding to the silicate/ice samples were 280 nm for $MgSiO_3$ and 800 nm for $H_2O$ (sample with the mass ratio of 0.8) and 180 nm for $MgSiO_3$ and 170 nm for $H_2O$ (sample with the mass ratio of 2.7).

Infrared spectra at four temperatures, namely 8 K, 100 K, 150 K, and 200 K, were measured in the transmission mode using a Fourier transform infrared (FTIR) spectrometer (Vertex 80v, Bruker) in the spectral range of 6000 – 200 cm$^{-1}$ with a resolution of 0.5 cm$^{-1}$, but will be presented further in the 4500 – 400 cm$^{-1}$ range because of the noisy low and high frequency edges. Due to slight instrumental instabilities of the baseline of the transmission spectra and possible multiple reflections on the silicate/ice layers, there can be artefacts affecting *k*-values in the ranges, where the *k*-values are supposed to be close to zero. Each higher temperature (100 K, 150 K, and 200 K) was achieved by heating the sample with the rate of 1 K min$^{-1}$. Before taking spectra at 100 K, 150 K, and 200 K the samples were allowed to stabilize for 30 min. The spectra of CsI substrates recorded before depositions at 8 K were used as reference spectra.

### 3. Data proceeding and calculation of optical constants

There are two sets of quantities that are typically used to describe the optical properties, the real and imaginary parts of the complex refractive index $n_1 = n + ik$ and the real and imaginary parts of the complex dielectric function (or relative permittivity) $\varepsilon = \varepsilon' + i\varepsilon''$ (Bohren & Huffman 2004). The relations between these two sets are the following:

$$\varepsilon' = n^2 - k^2, \ \varepsilon'' = 2nk \tag{1}$$

Two sets of the *n* and *k* constants have been obtained, namely, "measured" constants calculated directly from the transmission spectra of the silicate/ice samples and "effective" constants calculated from the *n* and *k* values of pure silicate and ice layers using different mixing rules. In both cases the *n* and *k* values were determined from Kramers-Kronig analysis of the transmission spectra. The reflection and interference losses were taken into account using an iterative procedure described in (Hagen, et al. 1981):

$$\alpha \equiv 4\pi k v = \frac{1}{d}\left(-\ln\left(\frac{I}{I_0}\right) + \ln\left|\frac{t_{01}t_{12}/t_{02}}{1+r_{01}r_{12}e^{2ix}}\right|^2\right) \tag{2}$$

$$n(v) = n_1^o + \frac{1}{2\pi^2}\int_0^\infty \frac{\alpha(v')}{v'^2-v^2}dv' \tag{3}$$

where α is the absorption coefficient, $I/I_0$ is the transmission spectrum, $d$ is the thickness of the sample, $v$ is the wavenumber, $v'$ is a running wavenumber in the integral, $r_{pq}$ and $t_{pq}$ are the complex reflection and transmission coefficients of the pq boundary, $x = 2\pi v d n_1$ and $n^0{}_1$ is a seed value for the real part of the index of refraction. For water ice, $n_1^o$ is 1.29 for amorphous $H_2O$ ice (in our case 8 K and 100 K) and 1.32 for crystalline $H_2O$ ice (in our case 150 K and 200 K) (Mastrapa, et al. 2009). Assuming a starting value for $n_1$, the absorption coefficient and the imaginary part of the refractive index were calculated from the transmission spectrum according to eq. (2). Then, $n$, the real part of the refractive index, was calculated by taking the integral in eq. (3) over the measured range since the denominator in eq. (3) gives little weight to contributions to the integral which are far from $v$. With this new refractive index, $n_1 = n + ik$, we calculated the transmission spectrum and compared it with the measured one. The procedure was repeated until 0.1% agreement between the measured and calculated transmission spectra was achieved.

For the silicate and silicate/ice samples these calculations were complicated by the porosity introduced by the layer of $MgSiO_3$ particles. Silicate grains are known to be very porous, the porosity can be as high as 90% (Sabri, et al. 2014). Fortunately, from the interference structures in the IR spectra due to the standing waves arising in the sample we could estimate the total film thicknesses of the porous layers to amount of 1460 nm for the silicate sample and 2200 and 1200 nm for the silicate/ice samples with the mass ratios of 0.8 and 2.7, respectively. The thicknesses were determined from the distance between fringes in the spectra. Using these values, eq. (2) and (3) delivered optical constants for the porous layer. Then, we calculated the constants for the inclusions, $MgSiO_3$ or $MgSiO_3/H_2O$, in vacuum using the inverted Maxwell-Garnett formula (see below) and the inverted volume fraction of the inclusions. This procedure delivered $k$ and $n$ for the silicate and silicate/ice samples. In these computations, $n_1^o$ was chosen in a way that the real parts of the measured optical constants came close to the effective ones after the correction of the porosity.

The effective optical constants were calculated from the $k$ and $n$ spectra of the $MgSiO_3$ and $H_2O$ samples using different mixing rules (often called effective medium approaches) for a two component mixture composed of inclusions (silicate grains) embedded in a matrix (water ice). In the next section we present mainly the results of the application of the Maxwell Garnett mixing rule working with the following approximations: i) a mixture composed of inclusions embedded in an otherwise homogeneous matrix, ii) the inclusions are identical in composition but may be different in volume, iii) the inclusions are separated by distances greater than their characteristic size, iv) the inclusions are spherical and small compared to $\lambda$. In this case the dielectric function of the effective medium can be calculated by the following equation (Bohren & Huffman 2004):

$$\varepsilon_{av} = \varepsilon_m \left(1 + \frac{3f\left(\frac{\varepsilon_i - \varepsilon_m}{\varepsilon_i + 2\varepsilon_m}\right)}{1 - f\left(\frac{\varepsilon_i - \varepsilon_m}{\varepsilon_i + 2\varepsilon_m}\right)}\right) \qquad (4)$$

where $\varepsilon_{av}$ is average dielectric function, $\varepsilon_m$ and $\varepsilon_i$ are dielectric functions of matrix and inclusions, and $f$ is the volume fraction of inclusions. The effective $n$ and $k$ values were obtained from $\varepsilon_{av}$ using relation 1. The other mixing rules tried were Bruggeman, Lichtenecker and Looyenga (Maron & Maron 2008).

Thus, for each silicate/ice ratio two sets of optical constants have been obtained: measured constants, calculated from the transmission spectra of the corresponding silicate/ice sample, and effective constants, calculated by mixing the $n$ and $k$ values obtained from the transmission spectra of the $MgSiO_3$ and $H_2O$ samples. The measured constants will be published in the Heidelberg-Jena-St.Petersburg Database of Optical Constants (http://www.mpia.de/HJPDOC/) and can be used in astrochemical models and for the decoding of astronomical spectra.

## 4. Results

The chosen temperatures of 8 K, 100 K, and 150 K correspond to different structures of water ice, namely high density amorphous ice, low density amorphous ice and crystalline ice (Hagen, et al. 1981; Jenniskens et al. 1995). At a temperature of 200 K, pure water ice (deposited without silicate grains) is completely desorbed from the substrate.

The observed vibrational bands in the spectra are the following: 3300 cm$^{-1}$ involving symmetric and asymmetric $H_2O$ stretching vibrations, 2347 cm$^{-1}$ – $CO_2$ stretching, 2200 cm$^{-1}$ – $H_2O$ combination mode, 1640 cm$^{-1}$ – $H_2O$ bending, 1040 cm$^{-1}$ – Si-O stretching, 770 cm$^{-1}$ – $H_2O$ librational motion, and 520 cm$^{-1}$ – O-Si-O bending. As an example of the measured spectra, transmission spectra of the silicate/ice sample with the mass ratio of 0.8, of the pure $MgSiO_3$ sample, and a pure $H_2O$ ice sample at 8 K can be seen in Figure 1.

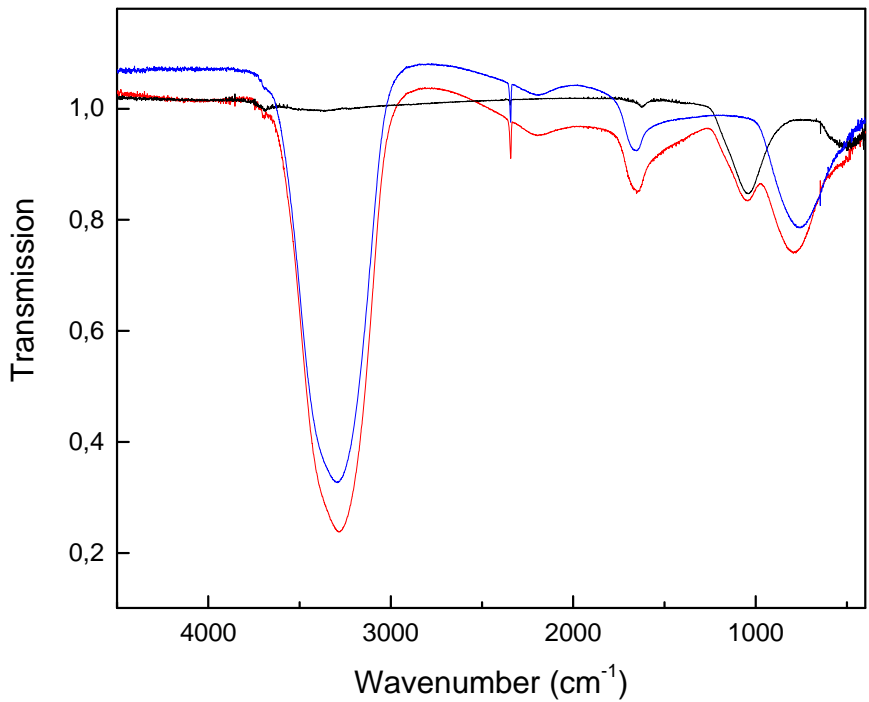

Figure 1. Transmission spectra of the $MgSiO_3/H_2O$ sample with the mass ratio of 0.8 (red curve), $MgSiO_3$ sample (black curve) and $H_2O$ sample (blue curve) at 8 K.

The optical constants of pure water ice and pure silicates obtained in this work were compared with the optical constants of amorphous and crystalline water ice (Mastrapa, et al. 2009) and amorphous, sol-gel MgSiO$_3$ (Jäger, et al. 2003) available in the Heidelberg-Jena-St.Petersburg Database of Optical Constants (http://www.mpia.de/HJPDOC/). There is a very good agreement between the optical constants of water ice derived in this study and the database ones. For silicates, the observed differences are not striking. The *k*- and *n*-values in the range of the stretching and bending vibrations of MgSiO$_3$ are comparable. Slight differences in the ranges, where the *k*-values are supposed to be close to zero are related to baseline instabilities and multi-reflection problem mentioned at the end of the experimental section.

The effective medium approaches discussed in the previous section were applied to silicate and water ice optical constants to obtain the effective optical constants for the silicate/ice mixtures. These effective constants were compared with the measured constants obtained directly from the transmission spectra of the silicate/ice samples. An example is presented in Figure 2. As one can see the mixing rules applied bring very similar spectra with small deviations in the regions of the water stretching vibration and librational motion and the silicate stretching vibration.

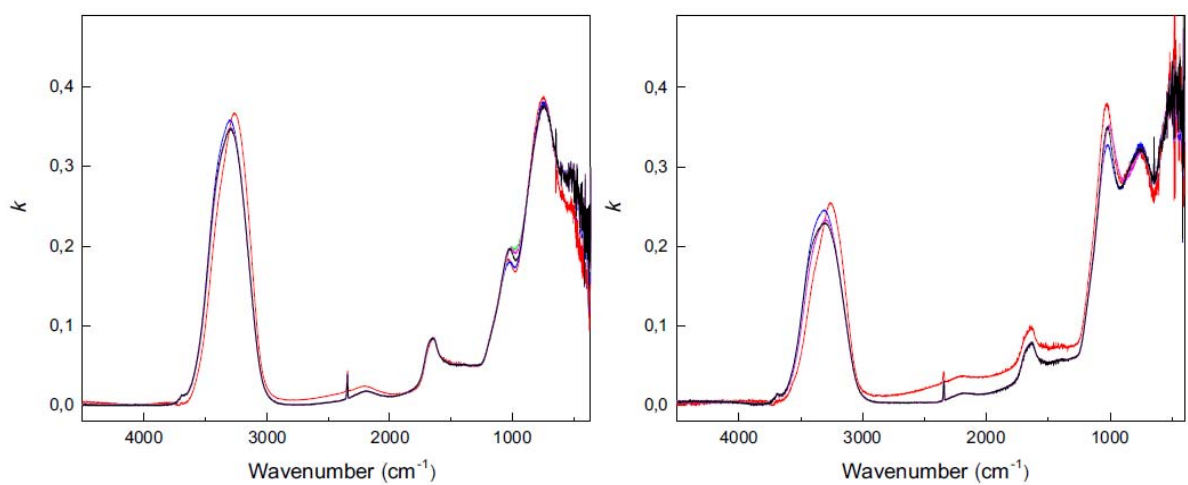

Figure 2. Measured (red curve) and effective *k* spectra at 8 K. Maxwell Garnett approach (black), Bruggeman approach (violet), Lichtenecker approach (blue), Looyenga approach

(green). MgSiO$_3$/H$_2$O mass ratio of 0.8 (left) and 2.7 (right). The effective curves practically coincide and the differences are visible only at enlargement of the figure.

The case of the silicate/ice mixture with the mass ratio of 2.7 is a bit more complicated because of the higher mass and volume fraction of silicates with respect to water ice. One can consider the sample as silicate inclusions in water ice matrix (as in the case of the silicate/ice sample with the mass ratio of 0.8) as well as water ice inclusions in silicate matrix. Bruggeman, Lichtenecker and Looyenga rules are symmetrical, i.e. the dielectric functions of matrix and inclusions can be exchanged. We have applied the asymmetrical Maxwell Garnett mixing rules for both silicate/ice mixtures. An example is shown in Figure 3. The differences between the spectra "silicates in ice" and "ice in silicates" are small, but a homogeneous water ice matrix is considered as more appropriate for the Maxwell Garnett approach, which requires a homogeneous matrix.

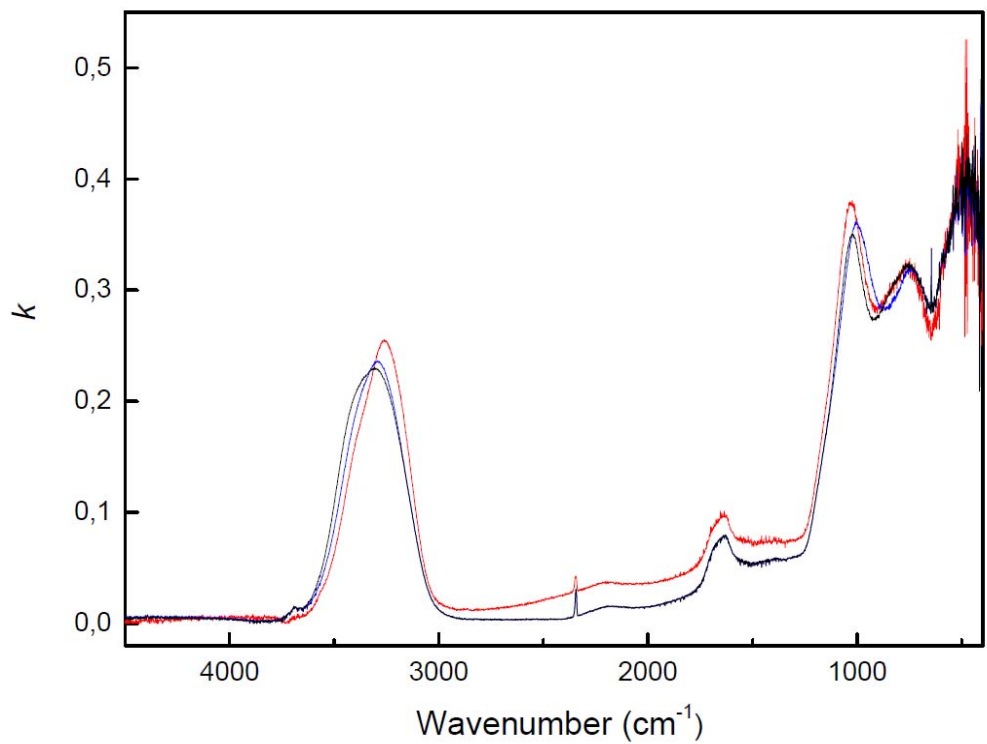

Figure 3. Measured (red curve) and effective $k$ spectra of the sample with the MgSiO$_3$/H$_2$O mass ratio of 2.7 at 8 K. Water ice as matrix (black), silicate grains as matrix (blue).

A comparison of the measured *k* and *n* constants of the silicate/ice samples with mass ratios of 0.8 and 2.7 with the effective *k* and *n* values is presented for different temperatures. The effective constants were obtained by using the Maxwell Garnett rule for the case of silicate grain inclusions in water ice matrix. Figure 4 presents the optical constants at 8 K for the two silicate/ice mixtures studied. A number of differences between the measured and effective spectra can be observed, namely a redshift, a slight narrowing, and an intensity increase of the stretching vibration of $H_2O$ in the measured spectra as compared to the effective ones. Also one can observe an increased absorption on the low frequency (long wavelength) side of the band, which is much more noticeable for the mixture with the mass ratios of 2.7.

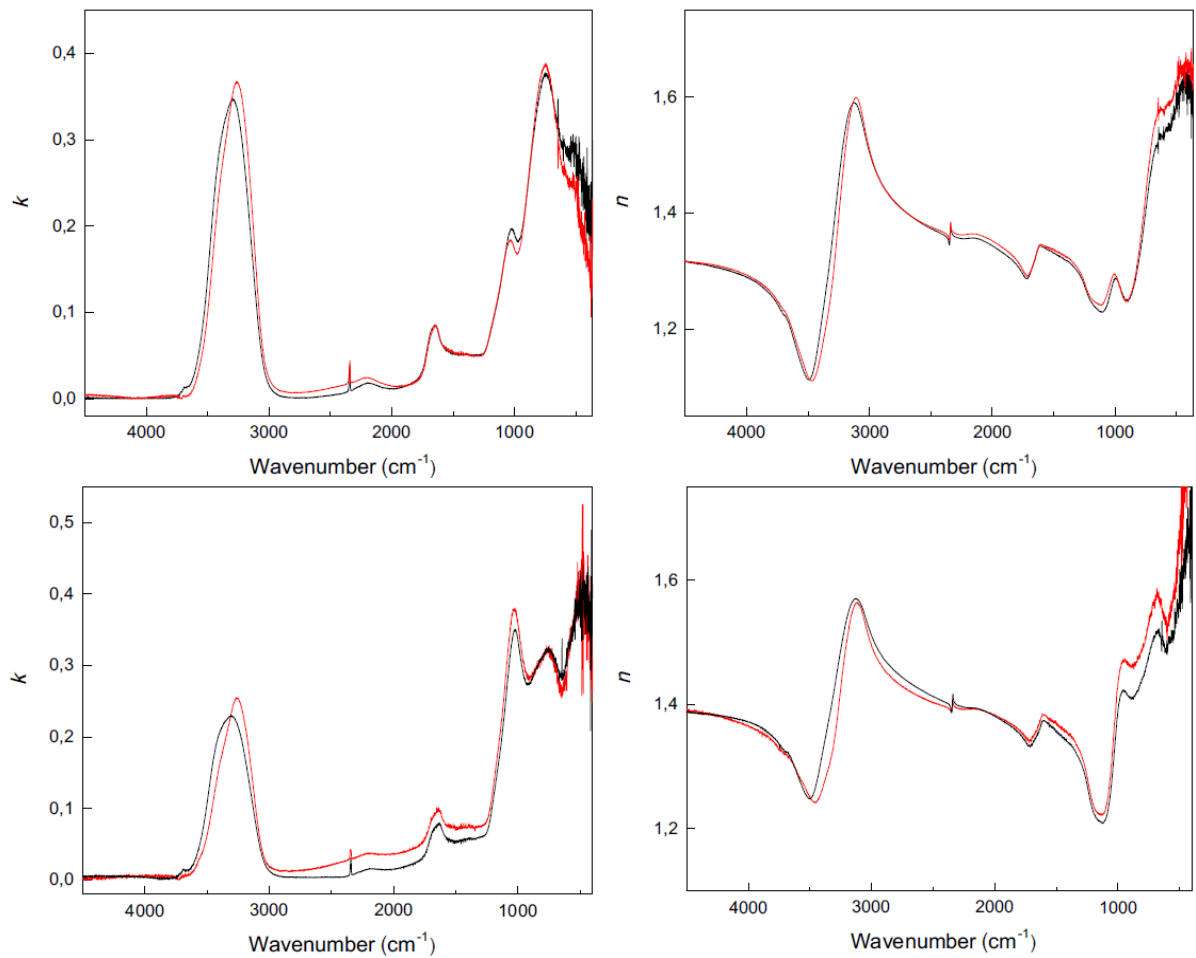

Figure 4. Effective (black curve) and measured (red curve) *k* and *n* spectra of the $MgSiO_3/H_2O$ mixtures at 8 K with the mass ratio of 0.8 (top) and 2.7 (bottom).

The first structural transformation of the ice from high to low density amorphous ice expected in the temperature range between 8 K and 100 K leads to the typical redshift, narrowing and intensity increase of the 3300 cm$^{-1}$ and 1640 cm$^{-1}$ water bands, a blueshift of the 2200 cm$^{-1}$ and 770 cm$^{-1}$ bands and to a decrease of the intensity of the stretching vibration of H$_2$O in the measured spectra with respect to the effective ones. This effect is again more noticeable for the sample with mass ratios of 2.7. The optical constants at 100 K are shown in Figure 5.

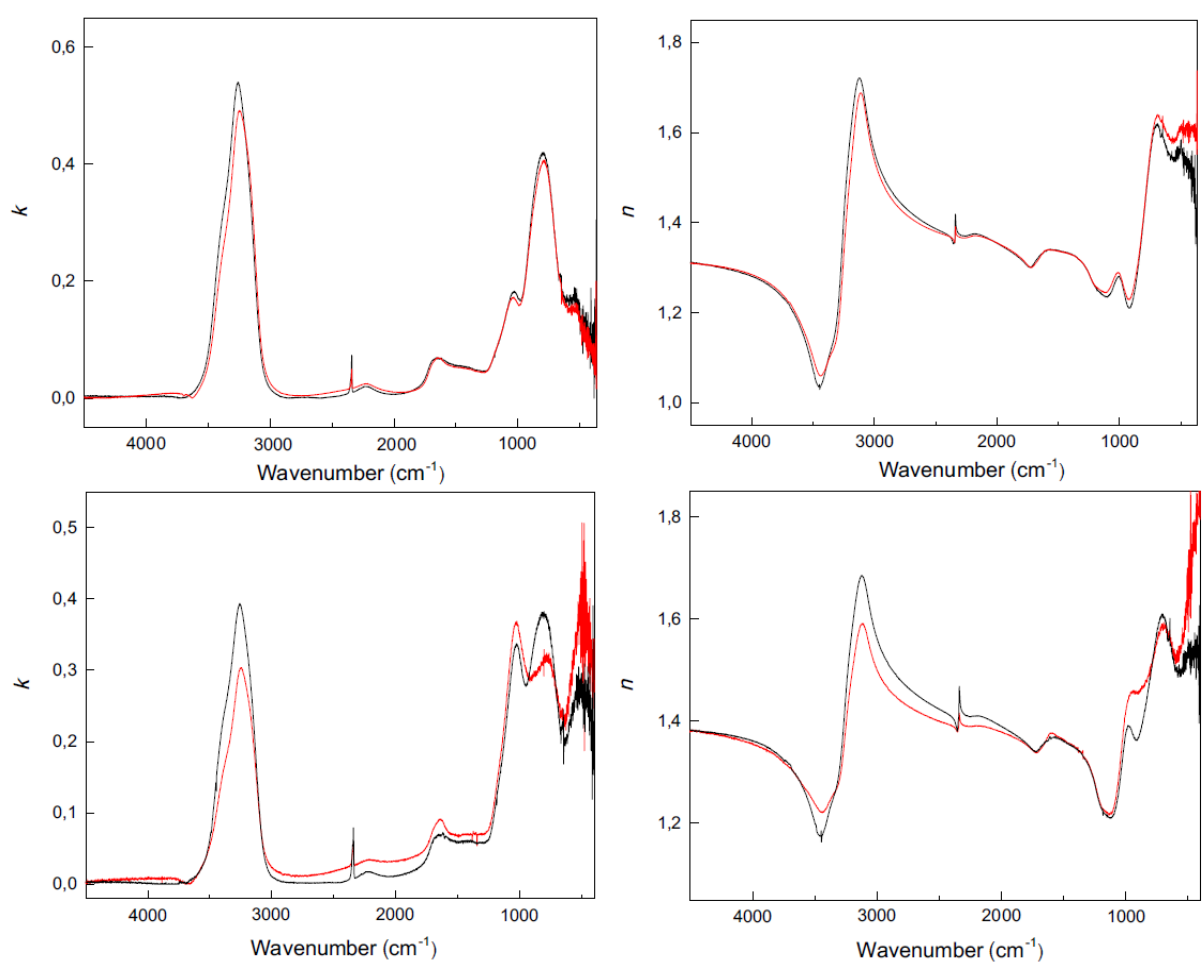

Figure 5. Effective (black curve) and measured (red curve) $k$ and $n$ spectra of the MgSiO$_3$/H$_2$O mixtures at 100 K with the mass ratio of 0.8 (top) and 2.7 (bottom).

The conversion of the amorphous into crystalline ice around 140 K leads to the typical redshift, narrowing, intensity increase, and appearance of a substructure in the 3300 cm$^{-1}$ water band. In addition, a suppression of the stretching vibration band of $H_2O$ in the measured spectra with respect to the effective ones can be observed. Optical constants of the silicate/ice mixtures at 150 K are presented in Figure 6.

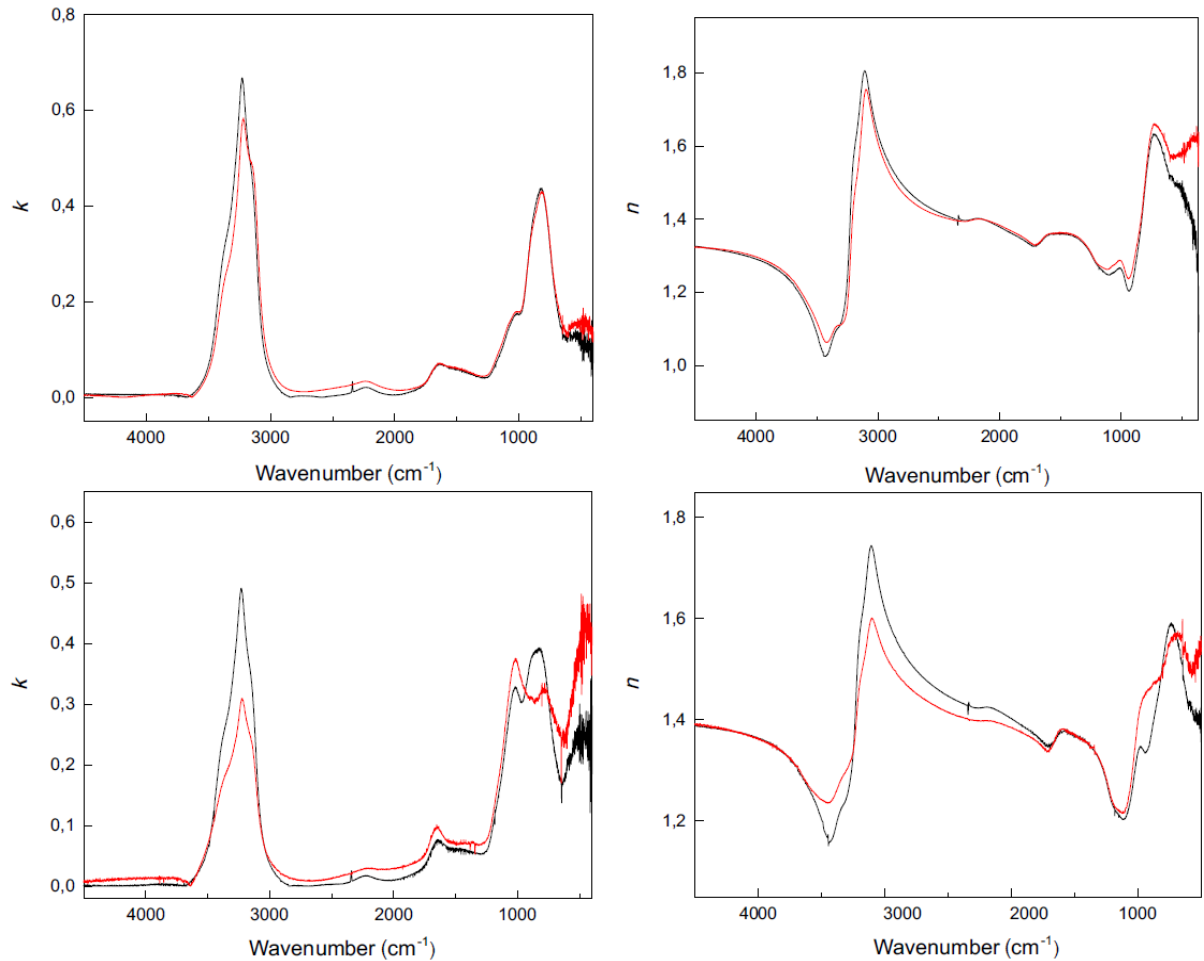

Figure 6. Effective (black curve) and measured (red curve) $k$ and $n$ spectra of the $MgSiO_3/H_2O$ mixtures at 150 K with the mass ratio of 0.8 (top) and 2.7 (bottom).

At 200 K we observe water absorption bands in the spectra of all silicate/ice samples while water ice in the pure $H_2O$ sample is completely desorbed. An example is presented in Figure 7.

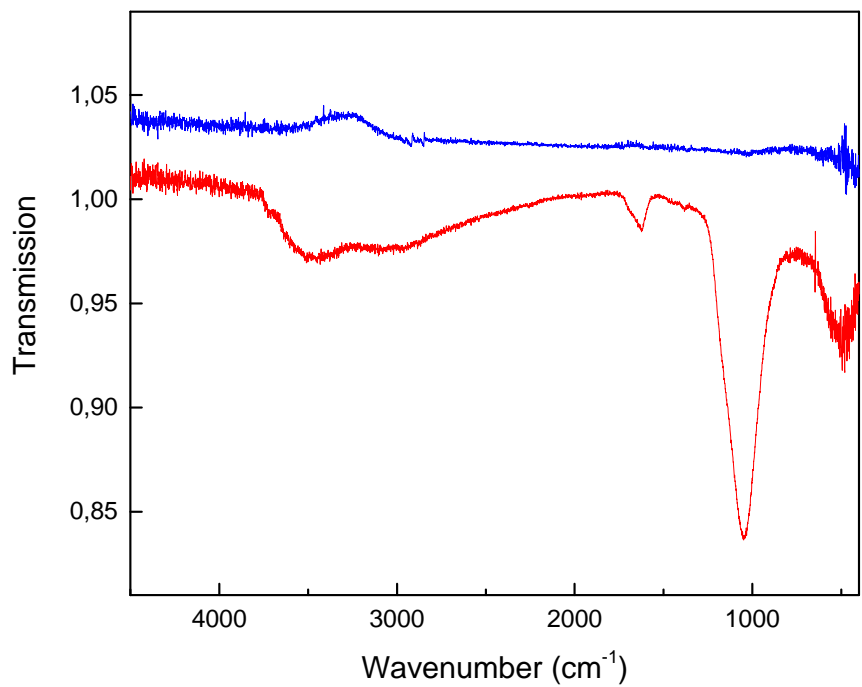

Figure 7. Transmission spectra at 200 K of the MgSiO$_3$/H$_2$O sample with the mass ratio of 0.8 (red curve) and H$_2$O sample (blue curve).

The water stretching band observed for the silicate/ice samples is very broad, which could be related to the morphology of the sample as well as to an insufficient baseline correction. A small positive bump in the water spectrum is due to the presence of a low amount of water ice in the reference sample (CsI substrate at 8 K before the deposition). The dependence of the relative (to deposited) remaining amount of water in the silicate/ice samples at a temperature of 200 K on the silicate/ice mass ratio is presented in Figure 8.

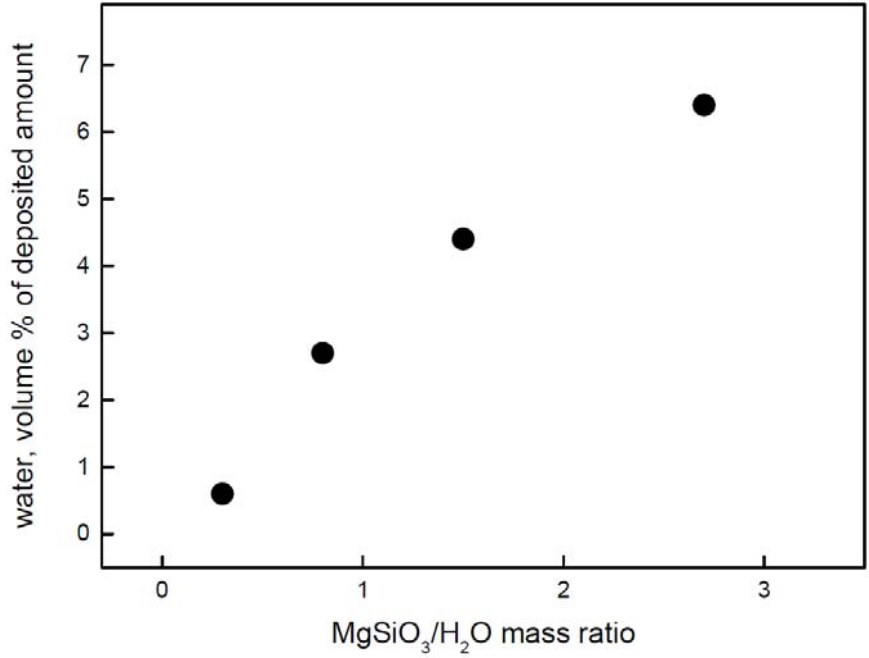

Figure 8. Dependence of the volume percentage of remaining water in the $MgSiO_3/H_2O$ samples at 200 K on the $MgSiO_3/H_2O$ mass ratio.

## 5. Discussion

With the approximations of the Maxwell Garnett mixing rule we can assume that the scattering on the grains is small and the extinction is the same as the absorption. In this case, a useful magnitude allowing to compare astronomical and laboratory measurements is the total absorption cross section (or absorption efficiency $Q_a$) of grains (Ossenkopf, Henning, & Mathis 1992). It can be taken per unit volume of grains or normalized to the particle radius. A comparison between the absorption efficiencies divided by particle radius, $Q_a/a$, calculated for small spherical particles in the Rayleigh limit from the measured optical constants is presented in Figure 9. $Q_a$ was defined as follows (Fabian et al. 2001):

$$Q_a = 2kV Im\left(\frac{\varepsilon}{\varepsilon - \varepsilon_m} \ln \frac{\varepsilon}{\varepsilon_m}\right) \qquad (5)$$

where $k$ is the wave vector, $V$ is the volume, $\varepsilon$ is the measured complex dielectric function, and $\varepsilon_m$ is the dielectric function of vacuum.

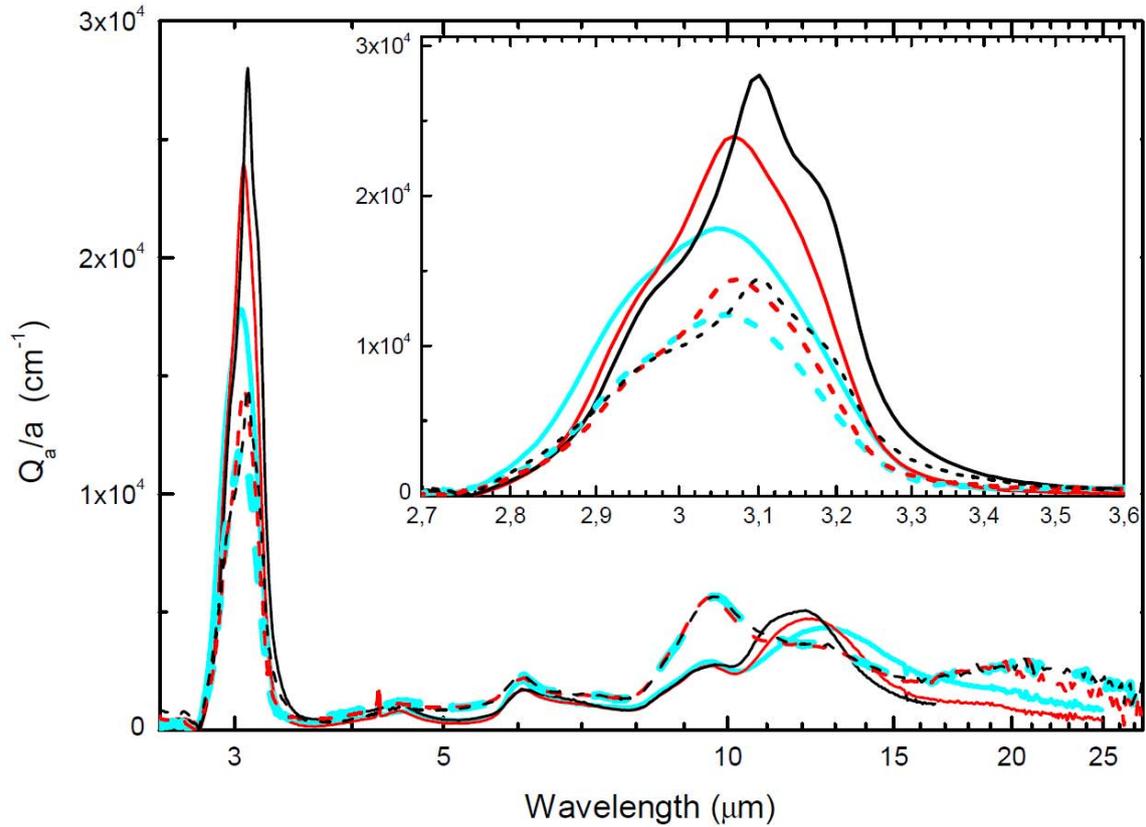

Figure 9: Normalized absorption efficiencies of silicate/ice composites with the mass ratios 0.8 (solid lines) and 2.7 (dashed lines) for 8 K (cyan), 100 K (red), and 150 K (black) calculated from the measured optical constants for a continuous distribution of ellipsoids.

For the two different silicate/ice composites with dust fractions of 0.8 and 2.7, the ratio between the OH stretching band at ~3 μm and the Si-O stretching band of silicates at about 9.7 μm is clearly changing. With increasing temperature, the sample characterized by the lower dust to ice ratio of 0.8 shows distinct shifts of the $H_2O$ stretching band at 3.1 μm to longer wavelengths and of the libration band at ~13 μm to shorter wavelengths. In addition, both bands are getting stronger. The position of the $H_2O$ stretching band at the maximum shifts from 3.05 μm at 8 K to 3.07 μm and 3.10 μm at 100 K and 150 K, respectively. In contrast, the libration bands show a short-wavelength drift from 12.78 μm to 12.34 μm and 12.01 μm, for 8 K, 100 K, and 150 K, respectively. With a higher dust fraction of 2.7, the 3.1 μm band is shifted comparably due to the modification of the ice structure with the temperature whereas the libration mode is unaffected by the temperature rise. The stretching band position varies from 3.06 µm, to 3.07 µm, and 3.10 µm for the temperatures of 8 K, 100 K, and 150 K respectively. Consequently, the temperature-related band shift of the $H_2O$ stretching band at around 3.1 µm seems to be independent of the silicate/ice ratio of the sample. The Si-O stretching vibrations shows only negligible changes with the temperature but a weak long wavelength shift of ~0.1 µm for the sample having a higher dust/ice ratio.

Experiments on various deposition rates of water at 10 K demonstrated that a decrease in the deposition rate shifts the peak frequency to lower frequencies while the band width decreases (Hagen, et al. 1981). It was shown later in a number of experiments that a lower deposition rate leads to a denser ice (Berland et al. 1995; Jenniskens, et al. 1995). High condensation rates limit the time for lateral diffusion following adsorption, thus, producing a lower density film (Berland, et al. 1995). Summarizing the results of these studies, we can conclude that the denser the deposited ice is the lower the band frequency is and the narrower the band is. Thus, in our case, the observed redshift and narrowing of the measured spectra can be linked to the denser water ice in the case of simultaneous deposition of water and silicate grains. The transformation from high-density to low-density ice also shifts the peak frequency to lower frequencies and narrows the band. It was supposed that this transformation requires the breaking of one hydrogen bond, on average, per molecule in the high density ice structure (Jenniskens, et al. 1995), which leads to a strengthening of the rest hydrogen bonds. One more explanation for the observed redshift and narrowing of the measured spectra can be the strengthening of hydrogen bonds due to the interaction with silicate molecules.

The astronomically observed increased absorption on the long wavelength side of the $H_2O$ stretching band was partly explained by scattering of large water grains (hundreds of nanometres) (Boogert, et al. 2015). In our case, the scattering by nanometre-sized grains covered by water ice may contribute to the increased absorption observed. However, this effect is much less noticeable for the laboratory grains as compared to cosmic grains mentioned by Boogert et al., probably, due to their much lesser radius. As one can see from Figures 4 to 6, the effect is more noticeable for the sample with the silicate/ice mass ratio of 2.7, where the grains can be bigger due to a larger amount of the material.

The suppression of the water stretching vibrations can be explained by different morphology of pure water ice and ice mixed with silicate grains. The transformation of the ice structure from high density to low density ice below 100 K and from amorphous to crystalline ice below 150 K leads to intensity increase of the water stretching band. We assume that the ice in silicate/ice mixtures converts incompletely and there is a noticeable fraction of high density ice at 100 K and of amorphous ice at 150 K. A change of the hydrogen bond strength due to the interaction with silicates can also play a role in the observed phenomenon.

All mixing rules applied fit the measured constants with comparable accuracy. The effective optical constants of silicate/water ice mixtures can be calculated by mixing the optical constants of the individual components, such as silicate grains and water ice. However, the position, width, and intensity as well as the increased absorption at long wavelengths of the water stretching vibration of the silicate/ice samples are not reproduced by the effective spectra. Thus, the mixing of the optical data of the icy silicate grain components to describe the optical properties of grains in the temperature range between 8 and 200 K can lead to incorrect results and complicate the analysis of astronomical data.

At 200 K, the water ice in the sample should be completely desorbed, but residual water is clearly detected in all silicate/ice samples. This is probably due to interactions between water molecules and the silicate surface. It is possible that a part of $H_2O$ molecules is chemisorbed on the surface of grains. Obviously, only molecules from the first monolayer of water ice can have stronger bonds to silicate grains or chemisorb on the grain surface. Also, water molecules can be trapped in pores of silicate grains. From our results in Figure 8, we can conclude that the higher the silicate/ice mass ratio is the greater is the number of binding sites or pores allowing more water molecules stay bound/trapped in silicates at 200 K.

The trapping of water ice in/on silicate grains is very important with respect to the presence of icy grains in protoplanetary disks, where grains are generally taken to be a starting point towards the formation of planets. Ice coated grains are expected to stick together much more efficiently. The increase of the mass in solid material in the region where there is ice compared to where there is no ice has been estimated to range from a factor of 1.6 (Min et al. 2011) to a factor of 4.2 (Thommes & Duncan 2006). Crystalline water ice in protoplanetary disks was tentatively detected (McClure et al. 2012; McClure et al. 2015). Recently it was unambiguously detected in a disk around the Herbig star HD 142527 (Min, et al. 2016). It was shown there that the disk contains a large reservoir of water ice, comparable to the water ice abundance in the outer solar system, comets, and dense interstellar clouds. The silicate/water ice mass ratio in the disk was determined to be ~0.63. This ratio was reproduced by one of our silicate/ice samples with the $MgSiO_3/H_2O$ mass ratio of 0.8. Water ice trapped in/on silicate grains and, thus, survived during the transition from a dense molecular cloud to a protoplanetary disk can partly explain large amounts of $H_2O$ ice in disks around young stars.

## 6. Conclusions

We have presented new sets of experimental optical data of silicate/water ice mixtures with the $MgSiO_3/H_2O$ mass ratios of 0.8 and 2.7 in the mid-IR spectral region at temperatures of 8 K, 100 K, and 150 K. The optical constants were obtained from the transmission spectra of the silicate/ice mixtures and compared with the effective constants calculated from the optical constants of pure silicate grains and pure water ice using the Maxwell Garnett mixing rule. Differences between measured and effective constants demonstrate that the determination of optical properties of interstellar and circumstellar icy silicate grains by a mixing of optical constants of water ice and silicates can lead to incorrect results and underline the importance of further investigations of optical and structural properties of grain/ice mixtures. A trapping of a part of water ice in/on silicate grains at 200 K is detected that can improve our understanding of the abundance and desorption properties of interstellar and circumstellar ices and can help to develop a link between the dust/ice ratio in a cosmic body and the cosmic environment, from where the body originates. The new optical constants of silicate/water ice mixtures published here can be used in models describing molecular clouds and circumstellar shells and for predictions of observables.

**Acknowledgments**

This work was supported by the Deutsche Forschungsgemeinschaft grant FOR2285 (sub-projects P5 and P8).